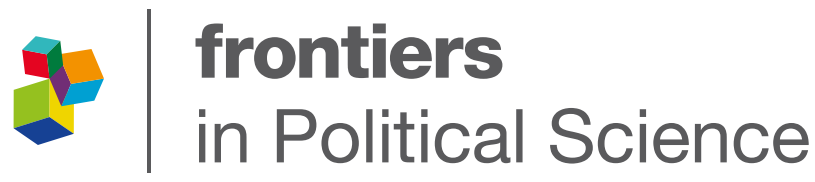

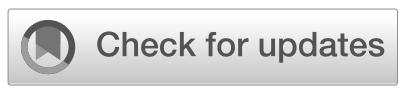

*CORRESPONDENCE
Ainara Larrondo-Ureta
✉ ainara.larrondo@ehu.eus

# Feminist voices, partisan networks: gender and political communication by MEPs on social media

Ainara Larrondo-Ureta[1]*, Simón Peña-Fernández[1], Julen Orbegozo-Terradillos[1] and Jordi Morales-i-Gras[2]

[1]Journalism Department, University of the Basque Country (EHU), Bilbao, Spain, [2]Cámarabilbao University Business School, Bilbao, Spain

This article analyses whether gender shapes Members of the European Parliament's communication on X/Twitter and whether equality-related and LGBTQ+ issues influence patterns of visibility, tone and interaction. Drawing on automated content, sentiment and network analysis of 122,972 tweets posted in 2022 by 620 MEPs, the study finds that women tweeted more frequently and were more visible in promotional content, while men favoured opinion-driven and news-focused messages. Equality and LGBTQ+ topics appeared more often in women's tweets (8.6%) than in men's (4.0%), and women's posts showed a more positive tone. However, interaction patterns were only marginally shaped by gender; instead, networks largely reflected political-group affiliation and national delegation. Overall, the findings suggest that although female MEPs engage more with equality-related issues, their digital communication remains embedded in established political structures, indicating gradual discursive change rather than substantive structural transformation.

## 1 Introduction

The political communication of parliamentarians extends beyond institutional procedures and formal deliberation (Coleman, 2006). It is a key arena through which representatives promote their political projects and engage with broader audiences (Lilleker and Koc-Michalska, 2013). Within this context, the European Parliament (EP), as a supranational and multilingual institution, represents a highly relevant empirical arena for exploring gendered political communication that transcends national and ideological boundaries.

This communicative environment also provides a useful lens through which to examine how broader debates on gender representation unfold within transnational political arenas. As women's presence in parliaments has expanded, it has generated expectations in areas such as the adoption of more inclusive agendas and the recalibration of policy priorities. Gender disparities continue to shape leadership distribution and internal party hierarchies, pointing to enduring structural asymmetries within representative institutions (Childs and Krook, 2009).

The EP offers a clear illustration. Over the past four decades, it has experienced a sustained increase in women's representation, rising from 15.2% in the first direct elections in 1979 to 39.3% in the current legislature (European Parliament, 2022). This trend has coincided with symbolic breakthroughs in EU leadership, including Roberta Metsola's appointment as President of the EP in 2022 and the 2019 appointments of Ursula von der Leyen and Christine Lagarde to the European Commission and the European Central Bank. Yet descriptive gains and high-profile appointments do not amount to substantive equality: women remain underrepresented and gendered asymmetries persist in access to influence and leadership within parliamentary structures (Shreeves and Boland, 2021; Abels et al., 2025).

Existing scholarship emphasises the need to scrutinise the norms and informal practices through which authority and influence are distributed within the institution (Ahrens and Rolandsen Agustín, 2019; Elomäki and Ahrens, 2022). In this regard, Kantola and Miller (2022), in their analysis of the eighth (2014–2019) and ninth (2019–2024) legislatures of the EP, demonstrate that men continue to dominate political leadership positions within parliamentary groups and national party delegations. By contrast, strategic leadership emerges as the most gender-balanced domain, although the European People's Party (EPP)—the largest and most conservative political group—constitutes a notable exception. Moreover, women remain underrepresented in the administrative leadership of political groups, further illustrating the persistence of gendered disparities within the Parliament's internal power structures.

These gendered disparities invite closer scrutiny of how they are reproduced—or potentially challenged—through communicative practices, particularly in the digital sphere and on platforms such as X/Twitter, which are widely recognised as central arenas for contemporary political discourse (Lappas et al., 2019).

At the same time, X/Twitter and other social media platforms do not constitute neutral environments for political communication. Platform logics and algorithmic dynamics shape political visibility on these networks, as well as forms of personalization and conflict (Van Dijck and Poell, 2013). In the case of women politicians, while these platforms may expand opportunities for public visibility and projection, they may also increase exposure to harassment, misogyny and other forms of gendered scrutiny (Harmer and Southern, 2021). In this sense, digital political communication should be understood as a practice that goes beyond individual preferences, reflecting communicative environments that may involve different forms of risk, disorder and inequality.

The use of social media by female candidates can enhance their visibility and increase coverage in traditional media, a factor associated with improved electoral prospects. At the same time, platforms such as X/Twitter may provide women politicians with a direct channel to reach the public and partially circumvent sexist filters, although these environments remain far from hospitable (Shreeves and Boland, 2021).

Politicians use X/Twitter within the logic of the permanent campaign, signaling responsiveness and sustained engagement with the public (Larsson, 2015). In the specific context of the EP, research has further shown that MEPs employ the platform to reach international audiences and to position themselves within a supranational political arena (Haman et al., 2023). Complementing this perspective, empirical studies have analysed patterns of participation through indicators such as account activity, frequency of posting and forms of interaction, including retweets, mentions and conversational exchanges (Scherpereel et al., 2017).

While this body of work has significantly advanced our understanding of how MEPs use social media, it has tended to privilege strategic communication logic over explicitly gendered analyses. To date, however, no study has systematically examined the differentiated use of social media by male and female MEPs from an explicitly gendered and feminist perspective—understood here as the public articulation of support for gender equality, women's rights, and the rights of marginalised gender identities (Kantola and Rolandsen-Agustín, 2019; Kantola and Lombardo, 2021).

This study addresses that gap by analysing tweet typologies, topic emphasis, sentiment, and network dynamics to assess whether gender structures digital visibility and whether feminist positioning translates into distinctive interaction patterns (Larsson, 2015). It contributes to research on gender and digital political communication in three ways: first, by providing a large-scale comparative analysis of multilingual parliamentary communication on X/Twitter; second, by integrating content, sentiment, and network analysis within a unified analytical framework; and third, by advancing a gender-sensitive perspective on political communication in supranational parliamentary contexts.

Gender is operationalised in binary terms for analytical purposes based on publicly available information (male/female). This decision reflects data availability rather than a dichotomous conceptualisation of gender; rather, gender is treated as a social and political construct shaping expectations, interaction and power relations (Childs and Krook, 2009).

Building on this understanding, the study adopts a gender-aware lens in its analysis of digital communication, thereby providing a framework for examining broader processes of institutional change, including the feminization of political spaces. In doing so, it aligns with recent research initiatives such as the EUGenDem (2022) project, which analyses gendered policies and practices within the EP's political groups.

Drawing on the literature reviewed, this study conceptualises gender as a cross-cutting factor whose potential influence is examined across three interconnected dimensions of MEPs' digital political communication. First, gender may be associated with different patterns of communicative visibility and activity, reflected in distinct forms of presence and participation on X/Twitter. Second, it may be related to the thematic orientation of communication, particularly with regard to issues connected to equality and LGBTQ+ rights. Third, it may contribute to shaping patterns of interaction within parliamentary networks, although these dynamics are also conditioned by factors such as partisan affiliation and the national background of representatives. Together, these three dimensions constitute the analytical framework that guides the formulation of the research questions and the empirical analysis conducted in this study.

The study thus seeks to determine whether female MEPs communicate differently on X/Twitter in terms of activity levels and emotional tone, and whether any such differences are associated with feminist positioning. To this end, a comprehensive analysis was conducted of tweets posted by male and female MEPs during the ninth legislature (2019–2024), guided by four specific research questions:

> *RQ1.* How does gender shape patterns of digital visibility and communicative presence among MEPs on X/Twitter?
>
> *RQ2.* To what extent does gender influence the incorporation of feminist, gender-equality and LGBTQ+ issues into MEPs' digital political discourse?

> *RQ3.* Does gender shape the emotional register through which MEPs frame political issues on X/Twitter?
>
> *RQ4.* Does gender operate as a cross-cutting factor in structuring interaction among MEPs on X/Twitter, or are these dynamics primarily shaped by classical factors of political behavior, such as party affiliation and national delegation?

## 2 Literature review

The progressive feminization of politics has channelled research toward gendered political communication, with research focusing on both the treatment of female representatives and gendered patterns of political communication (Childs and Kittilson, 2016). Women politicians are more frequently evaluated through references to appearance, clothing or private lifme mthan through their political opinions and decisions (Shreeves and Boland, 2021). Online, they are also more likely to be targeted by abuse and to respond to it (Wagner et al., 2017). High-profile cases—such as the viral criticism directed at Finland's former Prime Minister for sharing personal content— illustrate gendered double standards in digital political life (Badham et al., 2024) and demonstrate that platforms can expand access to visibility while simultaneously reproducing structural inequality.

Accordingly, research on gender bias and stereotyping has prioritised social media, particularly X/Twitter (Meeks, 2016; Usher et al., 2018; Mertens et al., 2019; Saluja and Thilaka, 2021). This scholarship stresses the importance of strengthening women's political voices online (Koc-Michalska et al., 2021), where hashtags such as #WomenInPolitics and #FeministPolitics contribute to issue visibility.

At the same time, these environments enable sexist and misogynistic attacks that constrain women politicians' expressive capacity (Krook, 2020; Harmer and Southern, 2021). This pattern is corroborated by the Inter-Parliamentary Union's survey of women parliamentarians, which shows that online harassment is frequently triggered by positions taken on human rights, including the defence of women's rights (Inter-Parliamentary Union (IPU), 2016). Hostile digital environments may also shape the topics addressed, the forms of self-representation and, more broadly, the communicative strategies employed by women politicians on social media.

A further strand of research has examined how gender stereotypes shape expectations of political style. Women's communication is frequently coded as empathetic and less competitive, whereas men's is associated with assertiveness and authority. Such contrasting expectations feed into broader role incongruity dynamics, reinforcing a perceived mismatch between leadership and femininity (Eagly and Karau, 2002; Nee and De Maio, 2019). In turn, these normative frames shape evaluative judgements, contributing to the underestimation of women's competence in high-stakes political domains (Aaldering and Van der Pas, 2018).

Historically, emotion was excluded from the hyper-rational public sphere and associated with the "feminine" as a marker of irrationality (Janz, 2000). In response, earlier strategies encouraged women leaders to downplay warmth to conform to masculine models of authority (Atkinson, 1984). More recent scholarship, however, has challenged this assumption, arguing for the political relevance of affect and its role in democratic engagement (Thompson and Hoggett, 2012; Schneider, 2014; Schneider and Bos, 2014). Empirical evidence now suggests that women politicians increasingly adopt more emotional and personalised forms of communication to engage citizens and articulate political positions (Beltrán et al., 2021), a shift that is particularly evident on social media platforms (Mohammad and Yang, 2011).

At the same time, gender does not operate in isolation from other structural variables. In their analysis of MEPs' use of Facebook and Twitter during the seventh and eighth legislatures, Lappas et al. (2019) identify gender, political group and member state as relevant factors shaping digital engagement, and call for the inclusion of additional variables such as popularity and follower counts. Similarly, Elomäki and Ahrens (2022) demonstrate that the implementation of gender mainstreaming within the EP remains contingent on political-group control and committee dynamics, underscoring the interaction between gender and institutional power structures.

Country-level studies likewise provide consistent evidence of gendered patterns in digital political communication. In the United States, female candidates and representatives were more likely to address so-called "female issues" on Twitter (Evans and Clark, 2016; Evans, 2016). In Spain, gender differences have been identified in audience reach and message diffusion (Guerrero-Solé and Perales-García, 2021), while machine-learning analyses suggest that politicians reproduce gendered issue agendas online, with men focusing more on politics, sport, ideology and infrastructure, and women more on gender and social affairs (Beltrán et al., 2021). Recent evidence from the United Kingdom similarly points to gendered differences in both patterns of platform use and linguistic style among MPs on X/Twitter (Harmer and Southern, 2021; Bauer, 2015). Beyond Europe, Haman (2023) finds that female parliamentarians in Latin America tend to display higher levels of activity on the platform and adopt distinct strategies to strengthen their visibility and participation in public debate.

## 3 Materials and methods

### 3.1 Sampling procedure

The EP provides a suitable setting for examining gendered patterns in political communication, given its diversity in nationality, party affiliation, language and demographic composition (European Parliament, 2022).

To address the research questions, the study combined quantitative text analysis with social network analysis.

A dataset was first assembled including the name, member state, national party and European political group of each MEP, drawing on the EP's website and selected personal webpages. In 2022, following the United Kingdom's withdrawal from the EU, the EP comprised 705 MEPs, of whom 620 (87.9%) maintained identifiable X accounts forming the basis of the dataset (367 men and 253 women).

Tweets were retrieved on 19 November 2022 using Twarc 2.14.0 and the timeline command. The initial script queried 654 MEP X/Twitter handles. Of these, 621 returned at least one accessible tweet, and 620 accounts could be matched with complete parliamentary metadata and gender information. For each account, the procedure retrieved up to 3,200 of the most recent accessible tweets, together with follower and mention relations recorded during the period under analysis (Summers et al., 2023). Cyprus was not represented in the final dataset, as the only female Cypriot MEP at the time of collection, Eleni Stavrou, did not maintain an X/Twitter account. As with all Twitter/X API-based collections, the dataset is subject to standard retrieval limitations. Protected, suspended, deleted or renamed accounts, deleted tweets, and content unavailable through the API at the time of collection may not be represented. The corpus should therefore be interpreted as the set of accessible tweets returned by the collection procedure at the time of retrieval rather than as an exhaustive archive of all activity. Importantly, no account in the final analytical corpus reached the 3,200-tweet threshold; the maximum number of retrieved tweets for a single account was 1,151, indicating that truncation did not operate as an effective constraint in the analysed dataset.

The tweets of the final sample were filtered by language and posts in unidentifiable languages (including image-only or emoji-only content) were excluded. This yielded a final corpus of 122,972 tweets. To enable cross-linguistic analysis, all non-English tweets were translated into English using Google Translate in Google Sheets, which provides coverage for all official EU languages and the minority languages present in our corpus. The use of Google Translate was adopted as a pragmatic solution for large-scale cross-linguistic comparison in a highly multilingual corpus produced by MEPs from different member states. This procedure allowed all tweets to be analysed within a common linguistic space and reduced the risk of privileging languages for which more mature NLP resources are available. However, machine translation may affect politically sensitive meanings, especially irony, idiomatic expressions, culturally embedded references, gendered language, hashtags and terms related to feminism, equality and LGBTQ+ rights. These limitations are particularly relevant for sentiment analysis and keyword-based thematic detection. Accordingly, the results are interpreted as aggregate patterns of thematic salience and tonal orientation rather than as fine-grained discourse analysis, and the study does not infer individual endorsement, ideological nuance or rhetorical intent from translated texts alone. The equality-related variable should therefore be understood as an indicator of thematic salience rather than as a direct measure of feminist endorsement or ideological positioning.

The cross-linguistic translation of tweets into English may introduce semantic simplifications or distortions, particularly in idiomatic expressions and culturally specific references. While translation enables large-scale comparative analysis, sentiment classification and entity recognition may be affected by subtle linguistic shifts. For this reason, the study focuses primarily on aggregate patterns rather than fine-grained linguistic nuances.

## 3.2 Analytical procedures

The analysis combined two complementary strategies: automated classification of tweet types, topics and sentiment, and network analysis of interactions among MEPs on the platform (Grandjean, 2016).

In the first case, tweets were classified by genre using the X-GENRE classifier (Kuzman, 2022; Conneau et al., 2019), which assigns short texts (minimum 75 characters) to categories including Information/Explanation, News, Instruction, Opinion/Argumentation, Forum, Narrative, Legal Content, Promotion and Other.

As a complementary step, named entities were extracted using spaCy's en_core_web_md model (Honnibal and Montani, 2017), focusing on the following types of entities: organization (ORG), person (PERSON), geopolitical entity (GPE), ethnic groups (NORP), product (PRODUCT), event (EVENT), law (LAW), work of art (WORK_OF_ART), building (FAC), language (LANGUAGE), location (LOC) and quantity (QUANTITY). Automated genre and sentiment classification inevitably involve classification error. Although the models employed have been trained on political corpora, their performance may vary across languages, rhetorical styles and national political contexts.

To examine the content of the tweets, terms were compared across genders by filtering for those showing a normalised presence above a predefined threshold (60%) and excluding entities used to a similar extent by both groups.

In addition, to identify the presence of gender- and feminism-related content in MEPs' messages, a keyword-based detection strategy was implemented using Python libraries, including Pandas. In line with the study's objectives, the following terms were selected: feminism, women, gender, patriarchy, equality, intersectionality, feminist, sexism, misogyny, gay, lesbian, trans, LGBT, and GLBT. The search procedure relied on partial-string matching to capture lexical variants. For example, feminism also captured feminisms, and LGBT captured extended forms such as LGBTI and LGBTQ.

The identification of equality-related content relies on keyword detection and therefore captures thematic salience rather than normative endorsement. Tweets may include critical, neutral or contextual references to feminism and LGBTQ+ issues that are not distinguishable through automated detection alone.

Sentiment analysis was conducted on the full corpus of tweets, drawing on established approaches to political sentiment detection (Ansari et al., 2020). Classification was performed using the XLM-T-Sent-Politics model (Antypas et al., 2022), developed by the Cardiff NLP group and available via the Hugging Face Transformers library.[1] The model has been trained on political tweets from Greece, Spain and the United Kingdom and is specifically designed to detect sentiment in political communication on X/Twitter.

Gender differences across the three core variables—tweet genre, equality-related content and sentiment—were assessed using chi-square tests of independence. Given the size of the corpus, statistical significance is interpreted alongside effect sizes, which provide the primary basis for substantive inference. Accordingly, Cramer's V is reported to indicate the strength of association between categorical variables on a scale from 0 (no association) to 1 (perfect association). The findings should therefore be interpreted as patterns of association, rather than causal relationships.

For the second analytical strategy, network analysis, two complementary graphs were constructed: (a) a follower network capturing relatively stable structural connections, and (b) an interaction network based on mentions and retweets, directly related to gender and interaction dynamics. Follower ties reflect a snapshot at the time of collection, whereas interactions are observed within the, 2022 window.

The analysis considered gender alongside other relevant variables available in the dataset, including member state, European political group, national party affiliation and age group.

1 https://huggingface.co/docs/transformers/v4.19.0/es/index

To detect community structures, the Louvain multilevel algorithm was implemented in Python. In addition, categorical assortativity coefficients were calculated using the NetworkX library (Hagberg et al., 2008)[2] in order to assess the extent to which connections were structured by shared attributes (Newman, 2002). In this framework, retweets and @-mentions generated edges in the interaction network, while follower relationships defined the structural network; these layers were analysed separately. These tools made it possible to determine which variables—such as gender, political group or country of origin—most strongly structured MEPs' interaction patterns, as reflected in their mention and follower networks.

Finally, assortativity was used to assess whether ties were primarily organised by gender or by other attributes such as political group, country of representation or national party affiliation.

Network visualisations were produced using Gephi (ForceAtlas2 algorithm) and complemented with Matplotlib outputs.

Together, these procedures provide a multi-layered framework for examining gendered patterns of expression and interaction among MEPs on X.

# 4 Results

## 4.1 Gendered patterns of digital activity

In aggregate, 620 Members of the EP (367 men and 253 women) posted 122,972 tweets over the course of, 2022. Male MEPs produced 68,856 tweets (56.0% of the total), whereas female MEPs produced 54,116 (44.0%). However, when activity is considered per account, women posted slightly more frequently than men, with a mean of 213.90 tweets per account compared to 187.62 among their male counterparts.

At member-state level, men generated more than half of the tweets in most countries, with notable exceptions in Spain (7,567 tweets by women; 6,196 by men), Finland (1,438 by women; 439 by men) and the Czech Republic (1,957 by women; 1,382 by men).

2 https://networkx.org/

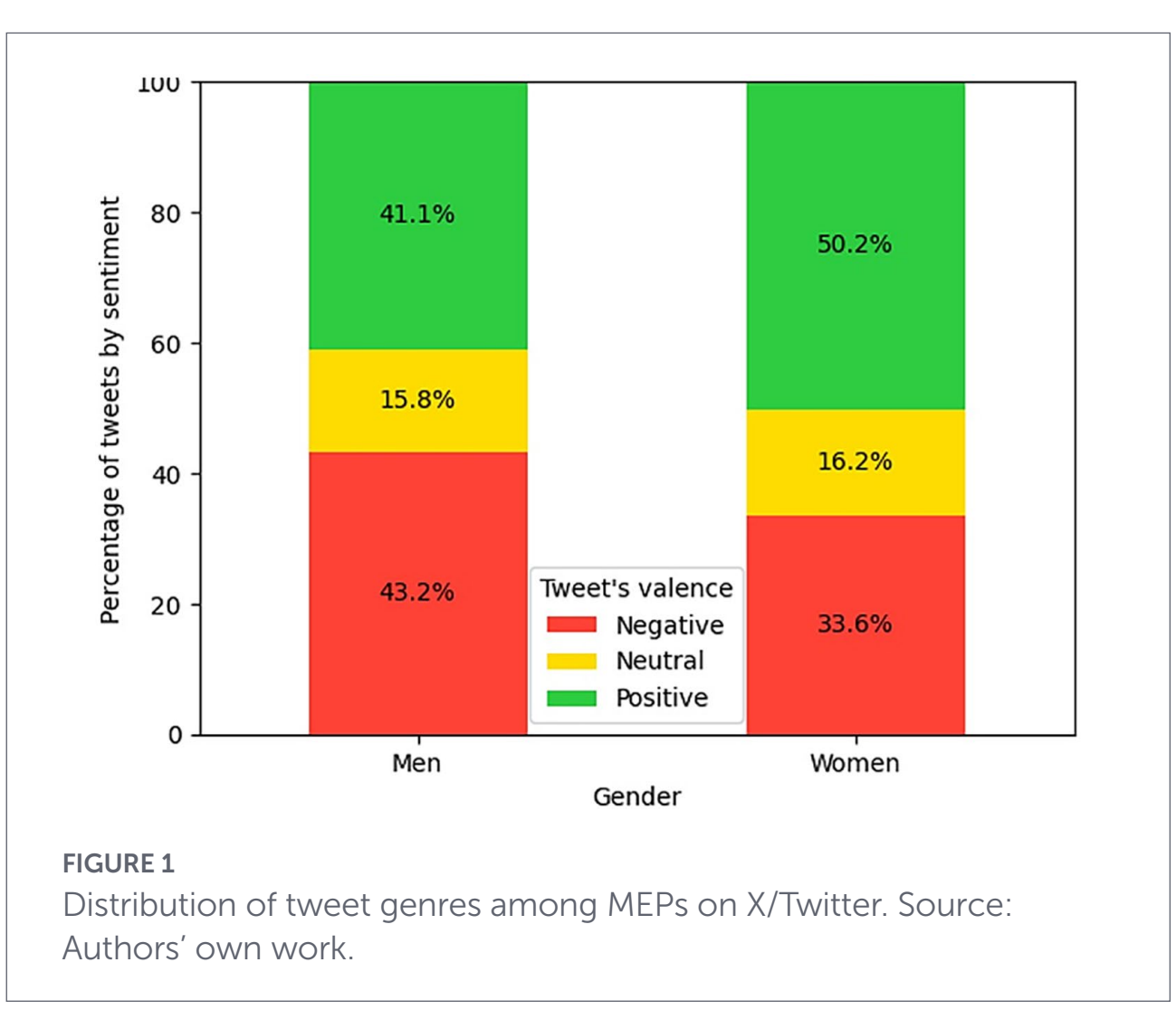

FIGURE 1
Distribution of tweet genres among MEPs on X/Twitter. Source: Authors' own work.

The results of the genre classification of tweets published by MEPs (Figure 1) show that the majority of posts (79,149; 64.36% of the total) can be classified as either Opinion/Argumentation or News, a distribution consistent with the political and informational logic of X/Twitter. At a considerable distance, first-person discussion and debate (Forum) and the promotion of events, initiatives or political activities (Promotion) also reached notable proportions, although clearly lower (13.23 and 10.04%, respectively).

Although genre classification indicates that Opinion/Argumentation and News dominate for both genders, the distribution across categories differs significantly across genders (chi-square, $p < 0.001$). Men are over-represented in Forum, Opinion/Argumentation and News, whereas women are over-represented in Promotion and Other.

Overall, women post slightly more per account, but occupy a more peripheral position within the hierarchy of genres (Promotion, Other), while men are proportionally more present in debate-oriented genres that shape the core discursive voice of parliamentary debate (Forum, Opinion/Argumentation and News). Although statistically significant, the strength of association is modest (Cramer's V = 0.098), and effect sizes vary across countries, reaching higher levels in Bulgaria (0.294), Latvia (0.271), Malta (0.269) and Greece (0.263), and remaining comparatively low in Italy (0.059), the Czech Republic (0.061), Belgium (0.078) and Slovakia (0.080), suggesting that gender interacts with broader contextual factors rather than acting as a dominant structuring variable.

## 4.2 Gender and equality discourse

The analysis of equality-related content in tweets published by MEPs on X/Twitter reveals a consistent gender gap in the articulation of feminist and LGBTQ+-related issues. Tweets were flagged through keyword detection covering terms associated with feminism, gender equality and sexual diversity. Across the corpus, 8.6% of tweets authored by women contained at least one equality-related term, compared to 4.0% of tweets authored by men (Figure 2). The difference is statistically significant ($p < 0.001$), although the magnitude of association remains modest (Cramer's V = 0.096). Gender is associated with a higher likelihood of addressing these issues, though it explains only a small share of variance in thematic emphasis.

Residual-based comparisons (observed vs. expected values) show that women are over-represented and men underrepresented in equality-related tweeting across most member states.

As shown in the lollipop graphs (Figure 3), women refer to equality-related themes more frequently than would be expected if gender were irrelevant, while men refer to them less frequently than expected. This pattern is visible in most member states, with exceptions in Croatia, Slovakia, Latvia and Malta.

In terms of the gap between observed and expected values, Austria and Lithuania display the largest deviations. However, in Lithuania the comparison must be treated with caution, as only one female MEP in the sample maintains an X/Twitter account—Aušra Maldeikienė, an independent member associated with the European People's Party—which limits the robustness of the national contrast.

When examining effect sizes through Cramer's V in those countries where the chi-square test yielded statistically significant results ($p < 0.05$; Table 1), notable cross-national variation emerges (Table 1). The association between gender and equality-related content is comparatively stronger in Austria (0.267), Ireland (0.153) and Poland (0.131). At the other end of the spectrum, countries such as the Czech Republic (0.030), Denmark (0.037) and France (0.058) display very small effect sizes, despite reaching statistical significance.

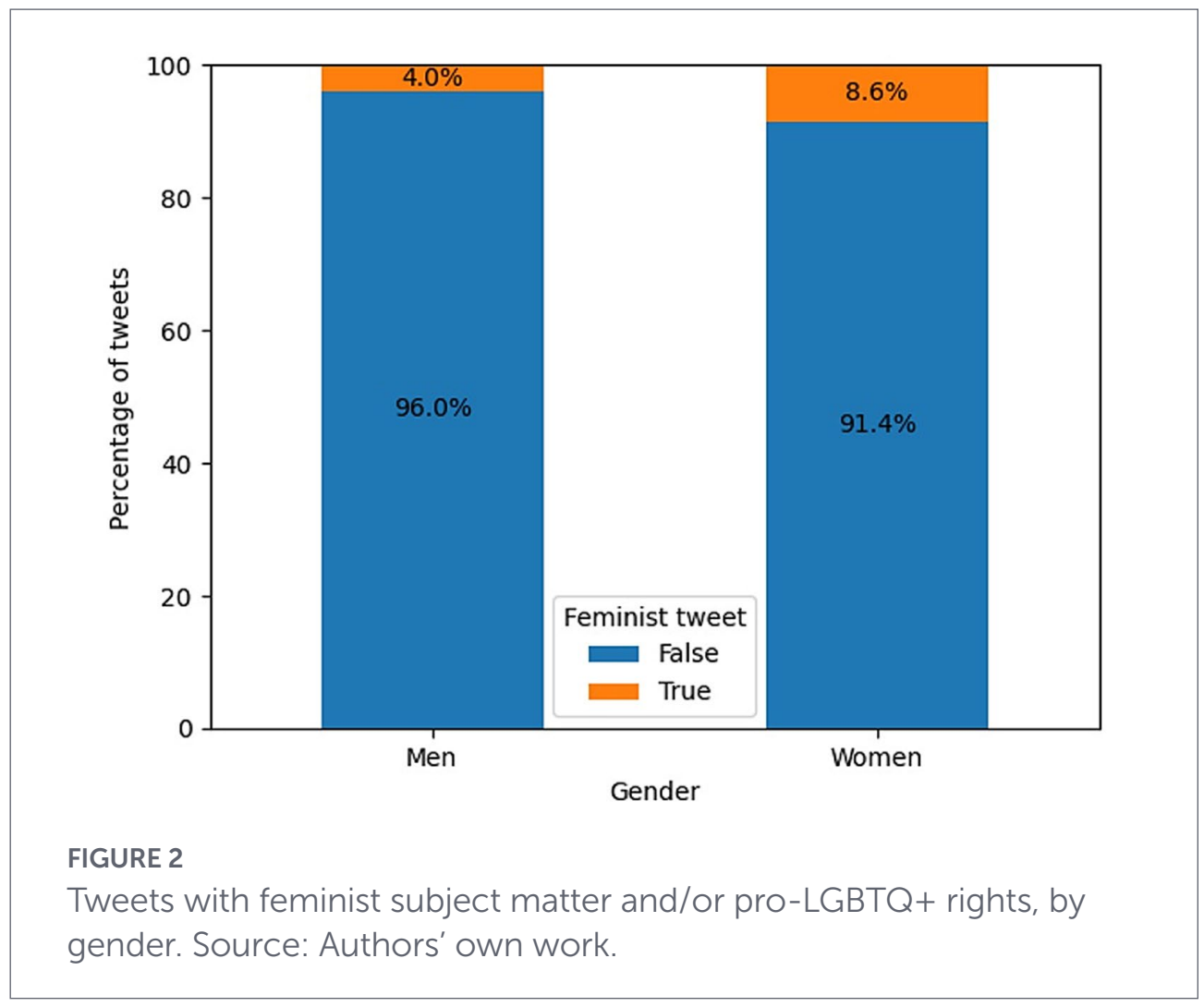

FIGURE 2
Tweets with feminist subject matter and/or pro-LGBTQ+ rights, by gender. Source: Authors' own work.

Cross-national differences should therefore be interpreted cautiously. The observed variation does not necessarily indicate stable national cultures of gendered communication, since country-level patterns may also reflect the composition of national delegations, party-system differences, the ideological distribution of MEPs, and the number of active accounts available in each member state. The results therefore suggest that the relationship between gender and equality-related discourse is context-dependent, but they should not be read as a ranking of member states. Rather, they point to the need to consider how gender intersects with national political dynamics and partisan composition. These variations may also reflect different national trajectories regarding gender equality, party competition and the politicisation of feminist issues across member states.

More broadly, these differences may also reflect variations in political culture, equality regimes and the persistence of gender stereotypes across national political contexts. Such factors may influence both the salience of equality-related issues and the incentives for political actors to engage with them in public communication.

Effect sizes vary across member states, suggesting that the gender–topic association is context-dependent. Given this variation, the analysis also considers other factors associated with equality-related tweeting.

On the one hand, ideological positioning also displays explanatory power in determining the extent to which content related to feminism and women is published. In this respect, the variable "Political Group" is also statistically significantly associated with the subject matter addressed in tweets ($p < 0.001$; Cramer's V = 0.076). MEPs affiliated with groups positioned further to the left in the Parliament—such as Renew Europe, the Progressive Alliance of Socialists and Democrats, the Left in the European Parliament (GUE/NGL) and the Greens/European Free Alliance—tweet about this subject matter to a greater extent than members of groups aligned to the right, including Identity and Democracy, the European Conservatives and Reformists, and the European People's Party. Within nearly all ideological families, however, female MEPs refer to equality-related issues more frequently than their male counterparts. The pattern is therefore not confined to a single political camp. It cuts across party families, albeit within the constraints of ideological positioning.

The age variable is likewise statistically associated with the thematic content of tweets, although the relationship is weak ($p = 0.001$; Cramer's V = 0.025). MEPs aged 20–30, 40–50, 50–60 and over 80 refer to feminism and LGBTQ+ rights more often than expected, whereas those aged 30–40 and 60–70 do so less often. The irregular distribution of these age groups, together with the small effect size, indicates that age has limited explanatory power in this case.

The analysis of thematic patterns beyond gender-related issues also indicates differences in the recurrence of topics according to the gender of parliamentarians (Figure 4). In the case of women, references are more evenly distributed and appear at relatively low percentages (approximately 0.3–0.9%). Mentions include party-related accounts such as @Renaissance_UE, national political contexts such as dkpol, collective actors such as Poles, and countries including Finland, Latvia and Afghanistan. Policy-related terms such as CohesionPolicy also appear, alongside diverse actors such as the Taliban and Amazon. The resulting profile is fragmented and multilayered, combining institutional, national and sectoral references without a single dominant focal point. By contrast, male MEPs display a markedly more concentrated agenda. Mentions reach substantially higher percentages (between 1.4 and 5.2%), with clear prominence given to Ukraine (5.2%), Russia (4.7%) and Putin (4.5%), followed by the US, Germany, China and NATO. The distribution shows a pronounced geopolitical axis structured around international conflict and major state actors.

These differences become clearer when considered in light of the exceptional geopolitical context of 2022. The Russian invasion of Ukraine made security, defence, sanctions, NATO and relations with Russia central topics in European political communication. The concentration of male MEPs' entities around Ukraine, Russia, Putin and NATO therefore reflects not only gendered thematic prioritisation, but also the dominance of a major geopolitical crisis in the parliamentary agenda. By contrast, the more fragmented distribution among women MEPs suggests a less concentrated agenda, combining institutional, national and policy-specific references. This does not necessarily indicate lower political salience, but rather a broader distribution of communicative focal points.

Overall, the contrast appears structural rather than incidental. Whereas male discourse clusters strongly around foreign policy and security issues, female discourse appears more diversified and less hierarchically concentrated. The data suggest different patterns of thematic prioritisation: greater geopolitical centrality among men and broader dispersion of references among women.

## 4.3 Gender and emotional tone

Gender differences in political communication also emerge at the level of emotional tone, although their magnitude remains limited. First, the results indicate a systematic but limited association between gender and sentiment. The chi-square significance ($p < 0.05$) shows that the distribution of sentiment categories differs by gender, but Cramer's V confirms that gender explains only a small share of the variance. Across the full corpus, tweets authored by women display a higher proportion of positive sentiment, while tweets authored by men show slightly higher levels of negative sentiment. The proportion of neutral tweets remains broadly similar across genders (Figure 5). In

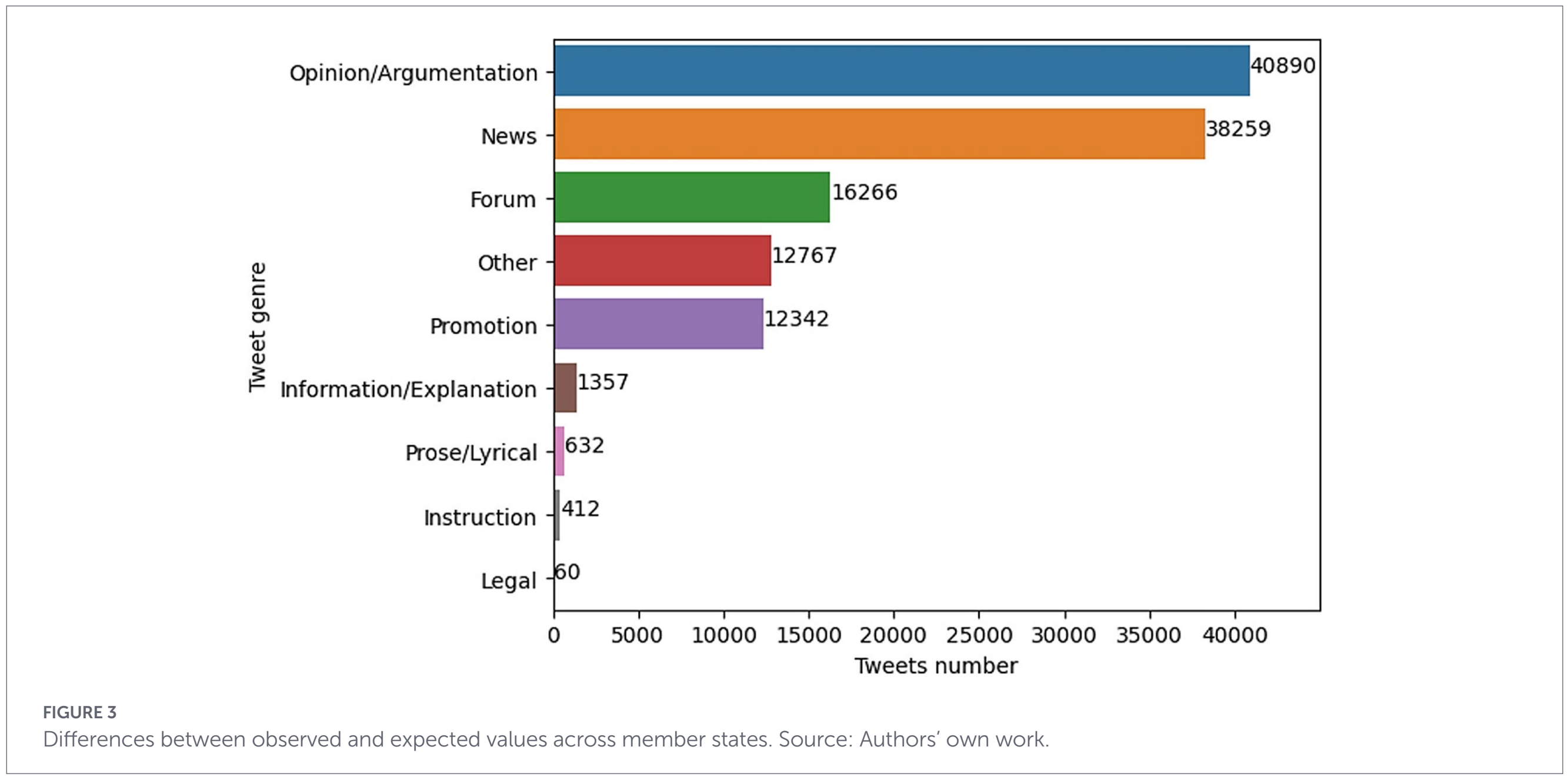

FIGURE 3
Differences between observed and expected values across member states. Source: Authors' own work.

TABLE 1 Effect size of the association between gender and subject matter for the countries producing a statistically significant Chi-squared test.

| Country | Cramer's V | Country | Cramer's V |
|---|---|---|---|
| Austria | 0.267404 | Netherlands | 0.094658 |
| Ireland | 0.153103 | Italy | 0.093857 |
| Poland | 0.131102 | Finland | 0.087097 |
| Spain | 0.127242 | Germany | 0.085201 |
| Bulgaria | 0.120822 | Portugal | 0.080731 |
| Lithuania | 0.108277 | Luxembourg | 0.068947 |
| Estonia | 0.106567 | France | 0.058302 |
| Sweden | 0.104374 | Denmark | 0.036613 |
| Belgium | 0.099538 | Czech Republic | 0.030595 |

Results for countries with non-significant chi-square tests are not interpreted.
Source: Authors' own work.

substantive terms, the difference is real, yet it does not suggest two separate communicative cultures; it suggests small shifts in tonal balance within a broadly shared repertoire.

When the analysis is restricted to equality-related tweets, the pattern persists. Messages posted by female MEPs on feminist and LGBTQ+-related issues tend more frequently to adopt a positive tone. Positivity rises among men from 41.1 to 45.5% and among women from 50.2 to 51.8%. This finding suggests that, within this corpus, feminism is not primarily framed through conflict or denunciation. Even among men—who tweet less about these issues—the tone of the posts that do address feminism moves in a positive direction.

The data indicate that the pattern cannot be reduced to a simple claim that "women are more positive"; rather, references to feminism are associated with a more positive emotional tone overall. When the analysis is restricted to equality-related tweets, the gender difference remains significant ($p < 0.001$), but the effect size is again modest (Cramer's V = 0.071). This suggests that sentiment differences persist even when topic is held constant, although their magnitude remains limited.

This suggests that the relationship between gender and sentiment should not be interpreted as a simple opposition between "positive women" and "negative men." Rather, equality-related discourse itself tends to be associated with a comparatively more positive or constructive register, while gender differences persist within this broader tonal pattern.

No member state reverses the overall tendency observed at the aggregate level. While country-level variation exists in the distribution of sentiment categories, the general pattern—greater relative positivity among women and comparable levels of neutrality—remains stable across national contexts.

To make the automated categories more interpretable, Table 2 provides anonymised paraphrases of tweets classified under the main analytical dimensions. These examples are not intended to be representative of the full corpus, but to illustrate the types of messages captured by the automated procedures.

## 4.4 Gender and network structure

Gender does not emerge as the primary structuring axis of interaction among MEPs; instead, network patterns are more strongly aligned with established political cleavages such as party affiliation and national delegation. Two complementary networks were analysed: a follower network capturing structural connections and an interaction network based on mentions and retweets.

In the follower network (Figure 6), the analysis focuses exclusively on connections among the MEPs included in the database, excluding mentions directed at external users. The resulting network is directed and weighted: each node represents an MEP's X/Twitter account and edges represent follower ties between MEP accounts.

Categorical assortativity reveals pronounced clustering by European political group and by country of representation. In substantive terms, MEPs tend to interact primarily with colleagues who belong to the same political group in the EP ($M = 0.543$), represent the same member state ($M = 0.399$) or belong to the same national party ($M = 0.204$). By contrast, gender has far weaker explanatory power in accounting for ties between MEPs ($M = 0.096$), as does age ($M = 0.068$).

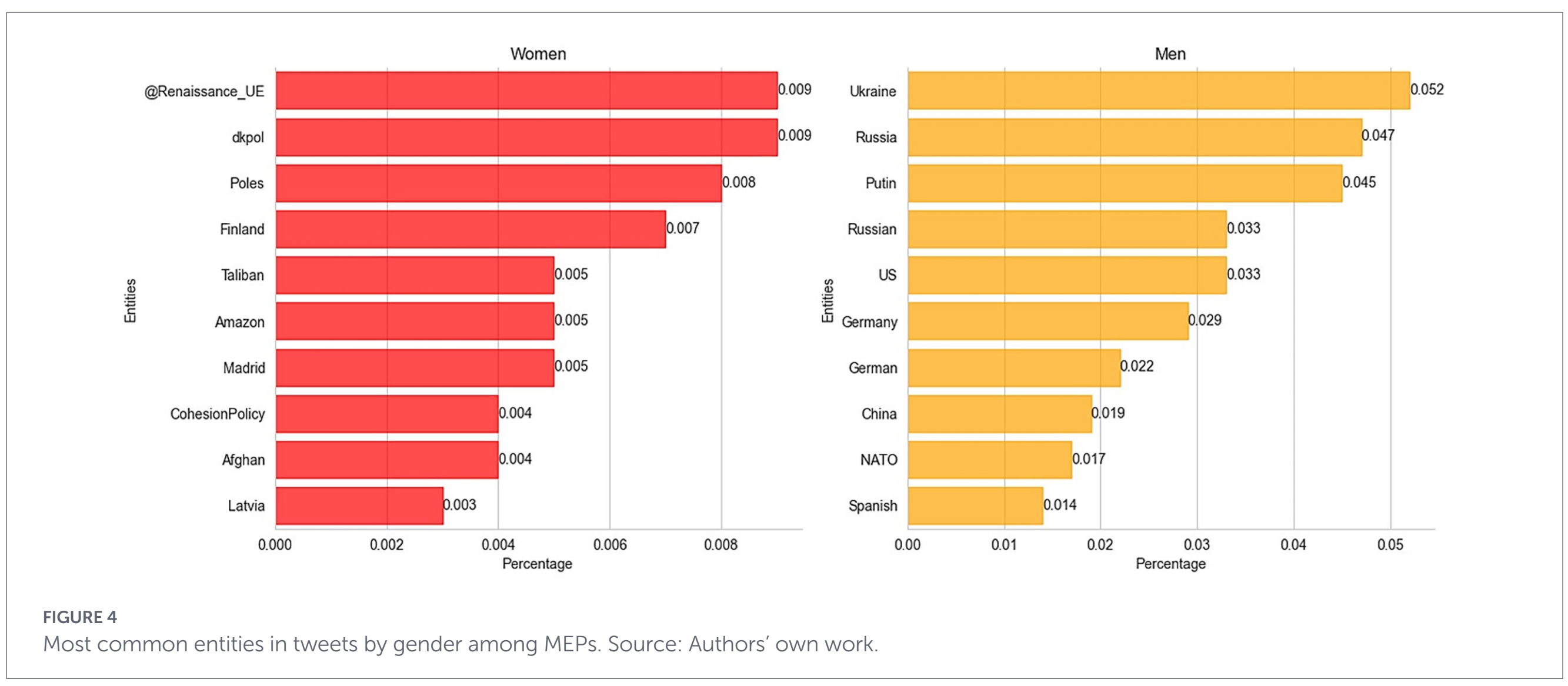

FIGURE 4
Most common entities in tweets by gender among MEPs. Source: Authors' own work.

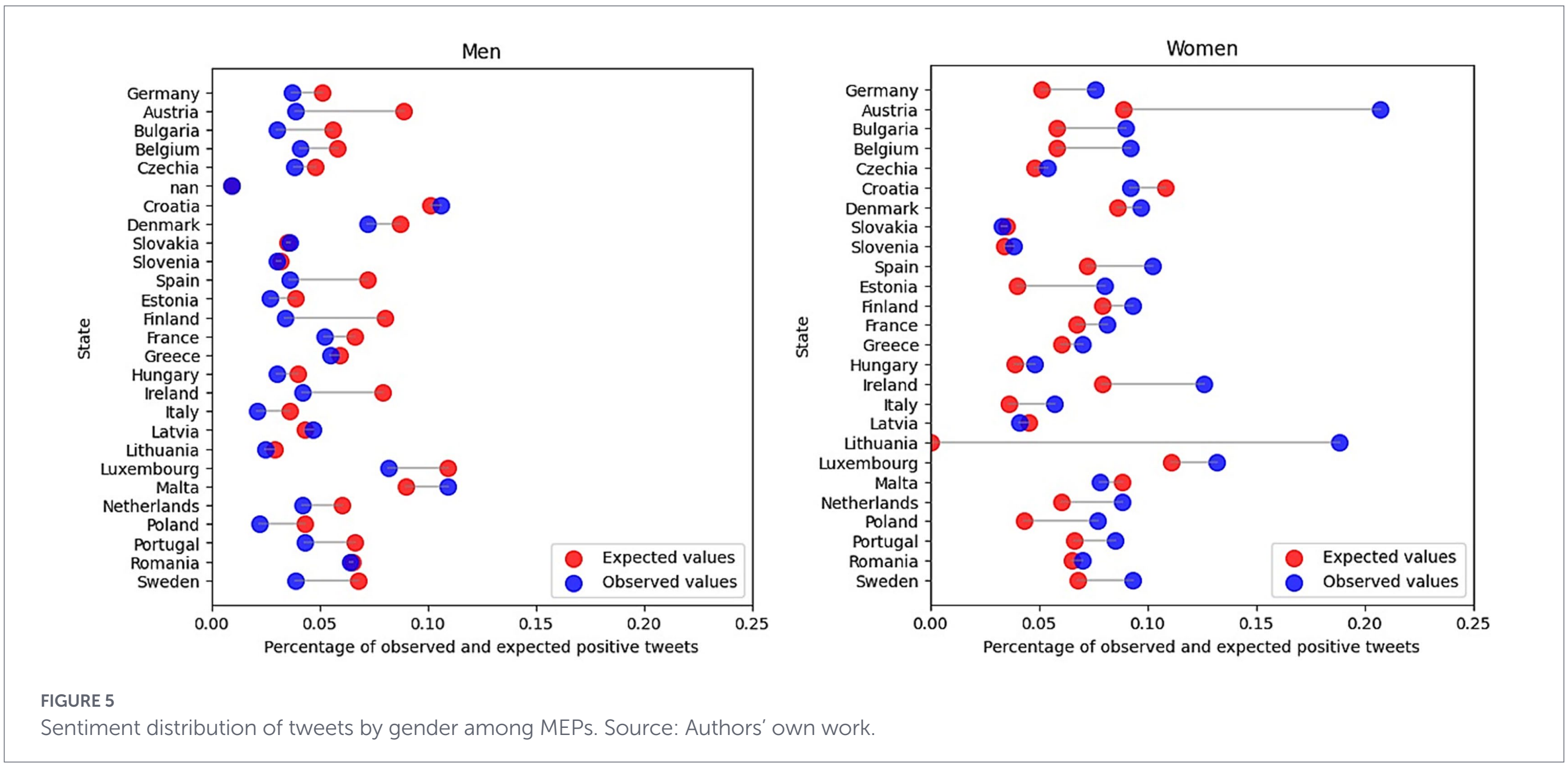

FIGURE 5
Sentiment distribution of tweets by gender among MEPs. Source: Authors' own work.

This distribution indicates that interaction in the EP's digital sphere follows the institutional logics that traditionally organise parliamentary behavior. The strongest connective principle is partisan alignment at European level, followed by national delegation, with national party acting as a secondary channel of cohesion. Gender, in turn, does not function as a cross-cutting basis for interaction that overrides these established affiliations. Rather, it operates—where it does at all—at the margins of a network primarily structured by political and territorial belonging.

The interaction network shows a similar configuration (Figure 7). Here again, the majority of MEPs predominantly follow colleagues from their own European political group ($M = 0.468$) and, to a lesser extent, representatives from the same member state ($M = 0.195$). As in the follower network, the interaction network also shows that the tendency to follow colleagues of the same gender is much weaker ($M = 0.033$), a pattern mirrored in the age variable ($M = 0.020$).

These figures reinforce the interpretation that digital connectivity within the Parliament remains structured primarily by European-level partisan alignment, with national delegation playing a secondary role. Gender and age exert only limited influence, confirming that they do not function as central organizing principles of digital affiliation.

## 5 Conclusion

This study set out to examine whether gender is associated with patterns in the digital communication of MEPs on X/Twitter (Scherpereel et al., 2017). Drawing on a large-scale analysis of tweet genres, thematic emphasis, sentiment and network interactions, the findings reveal subtle gendered differences in expression that coexist with interaction patterns largely structured by political group and national delegation.

First, activity levels reveal a balanced but differentiated pattern. Women MEPs' digital presence is as strong as men's, and they even

TABLE 2 Illustrative anonymised examples of tweet categories identified in the analysis.

| Analytical category | Anonymised paraphrased example |
|---|---|
| Equality-related discourse | A tweet welcomes progress on the Women on Boards Directive after years of deadlock, framing the measure as a major step towards gender equality in European companies. |
| Equality-related discourse with negative tone | A tweet criticises the withdrawal from or weakening of the Istanbul Convention, warning that women victims of domestic violence would be less protected. |
| Promotional style | A tweet announces or celebrates participation in an institutional event, meeting or campaign, inviting users to follow the activity and framing it as part of broader EU work. |
| Geopolitical discourse | A tweet condemns Russia's aggression against Ukraine and calls for stronger EU action, including sanctions or measures targeting Russian economic and political power. |
| Positive emotional tone | A tweet congratulates a colleague on a new institutional position and expresses enthusiasm about future cooperation for a fairer, greener and more feminist Europe. |
| Negative / opinion-driven communication | A tweet describes a political decision as shameful or unacceptable and calls for a stronger institutional response, using an explicitly evaluative and confrontational register. |

Source: Authors' own work.

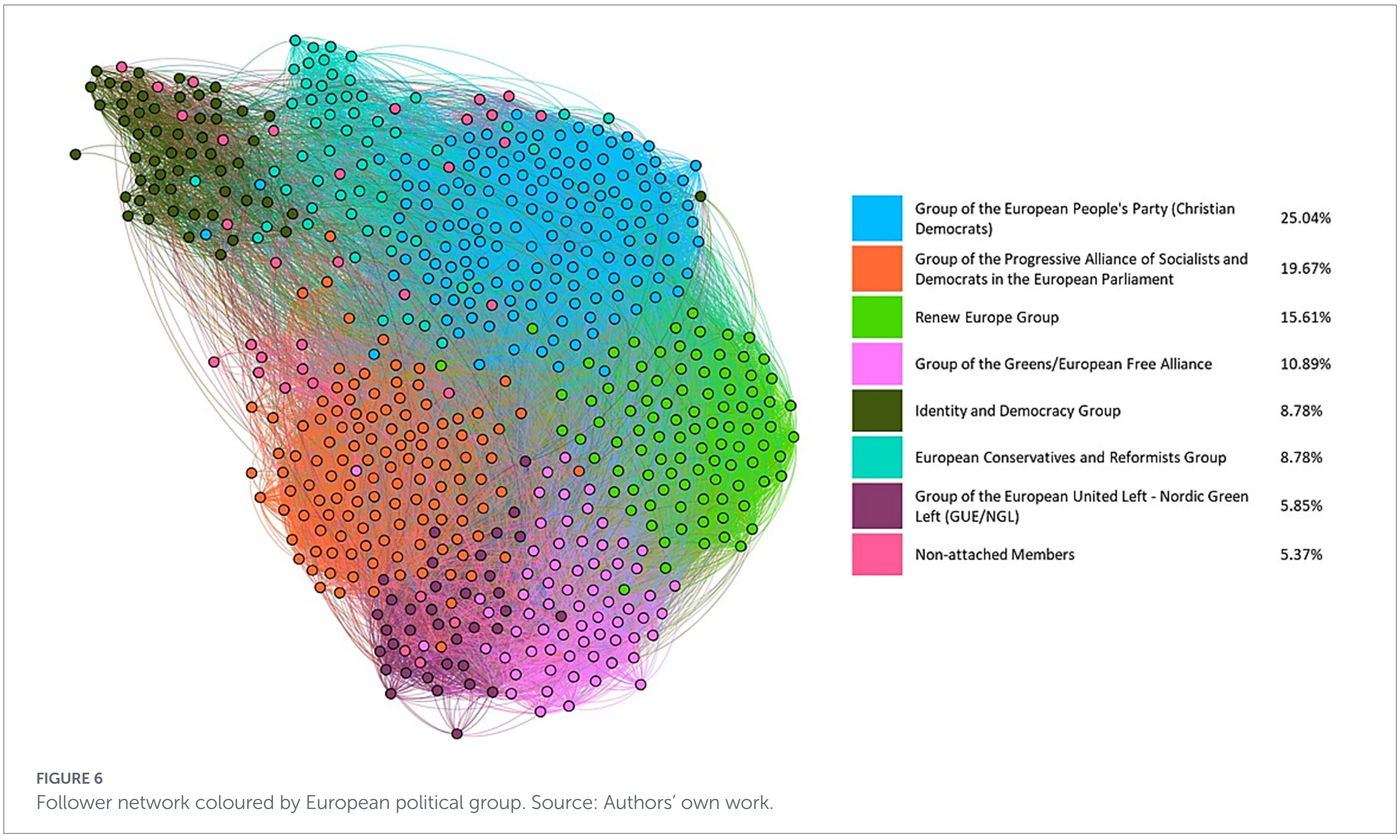

FIGURE 6
Follower network coloured by European political group. Source: Authors' own work.

post slightly more frequently on a per-account basis. However, genre distribution reveals qualitative differentiation. While both genders concentrate on posts that state political positions or report on current developments, men are proportionally more present in debate-oriented and discursive genres, whereas women are relatively more visible in promotional and residual formats.

Second, thematic emphasis shows a clearer gender gap. Women refer to feminism and LGBTQ+ topics approximately twice as often as men, although equality discourse remains a minority strand even within women MEP's communication. Political-group affiliation also shapes the visibility of this content: MEPs from centre-left groups address it more frequently than those aligned with the right. The findings suggest that gender differences in thematic salience unfold within clear partisan boundaries.

Third, sentiment analysis reveals tonal differentiation. Women's tweets are slightly more positive. When the analysis is restricted to feminism-related tweets, positivity increases for both groups. This pattern suggests that discourse on feminism and LGBTQ+ issues is not primarily articulated through conflictual or adversarial framing, and that content addressing women and feminism, while not uniform in tone, tends overall to adopt a comparatively more positive register.

Finally, network analysis shows that follower and interaction patterns are overwhelmingly structured by political group within the EP

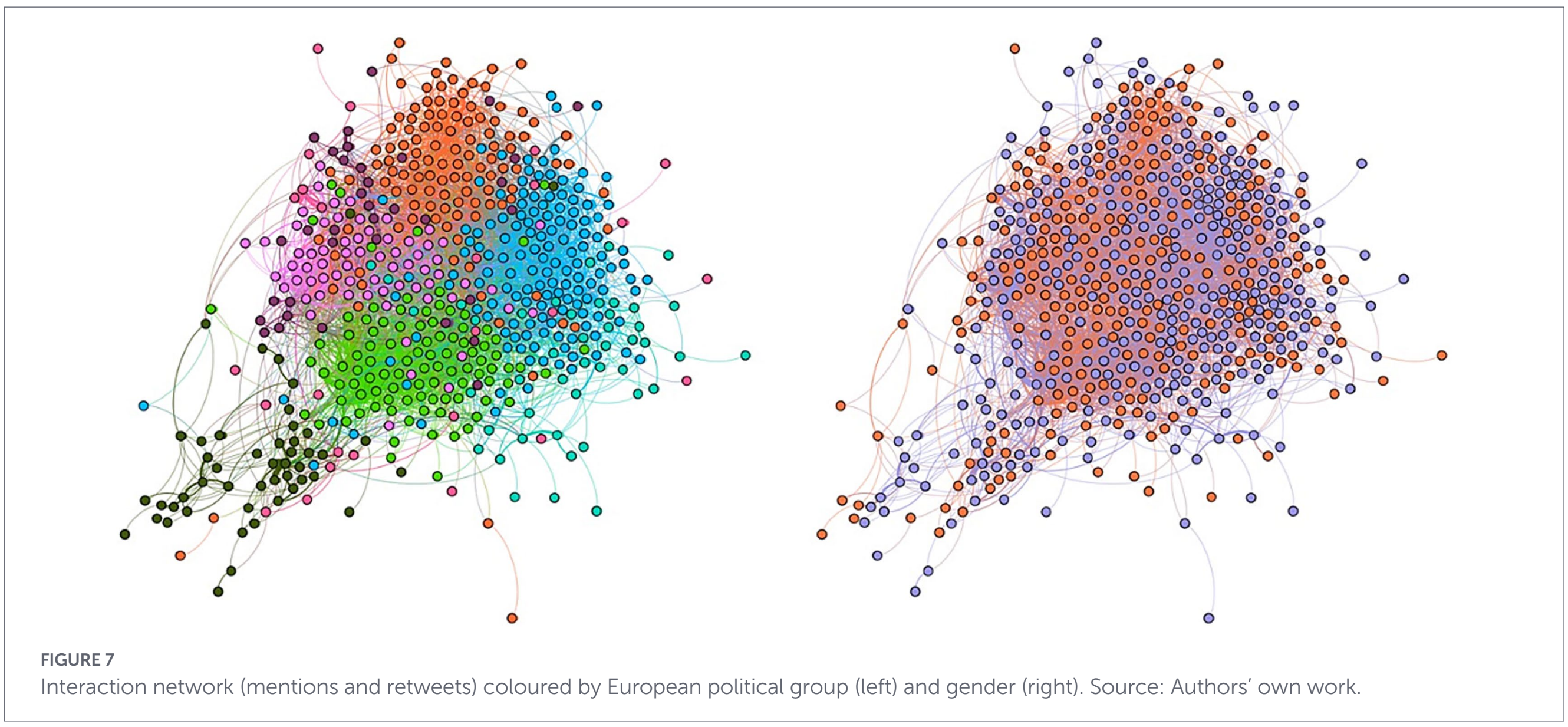

FIGURE 7
Interaction network (mentions and retweets) coloured by European political group (left) and gender (right). Source: Authors' own work.

and by the country represented. Assortativity coefficients for partisan alignment are substantially higher than those for gender. No gender-based clusters emerge, and gender does not function as a cross-cutting principle capable of reshaping digital connectivity.

The analysis also points to cases that do not fully conform to the dominant pattern. Some male MEPs explicitly engaged with equality-related and feminist discourse, while some female MEPs adopted strongly geopolitical or confrontational registers, particularly in relation to the war in Ukraine. These exceptions are analytically important because they show that gendered communication patterns are probabilistic rather than deterministic. Gender shapes the likelihood of certain thematic and tonal orientations, but it does not mechanically define individual communicative behavior.

# 6 Discussion

The central implication of these findings is that gendered differences in expression do not translate into cross-cutting coordination among MEPs on X/Twitter (Scherpereel et al., 2017). Feminist positioning appears primarily as a feature of content production—what is said, how often, and with what tonal register—rather than as a basis for alliance formation or relational clustering.

Online ties remain anchored in the Parliament's core organizational logics, shaped mainly by European political group and national delegation, with gender exerting only marginal influence on who interacts with whom (Guerrero-Solé and Perales-García, 2021; Mertens et al., 2019). This distinction matters analytically because it separates visibility and agenda articulation from the organization of political relationships: the former shows modest gendered differentiation, while the latter largely reproduces institutional structures in digital form (Kantola and Lombardo, 2021).

The results therefore reveal a clear tension: while X/Twitter appears to increase the visibility and dissemination of equality- and representation-related discourse, particularly among women MEPs, interaction dynamics continue to reproduce the partisan and national structures of the European Parliament.

In this sense, and in line with previous contributions on the logics of digital platforms (Dahlgren, 2005; Van Dijck and Poell, 2013), X/Twitter can be understood as a digital public sphere partially shaped both by algorithmic dynamics of visibility and by the political and institutional structures that organise parliamentary representation. This generates a complex interaction between discursive openness, political visibility and the reproduction of traditional power dynamics.

Ideological positioning further qualifies the role of gender in shaping digital discourse. Women refer more often to feminism and LGBTQ+ issues, although the overall presence of these themes remains relatively limited (Larrondo-Ureta et al., 2019). At the same time, ideological affiliation conditions their visibility more strongly than gender alone. Equality-related discourse is more common in center-left groups, regardless of gender. This suggests that feminist or equality-oriented content depends not only on descriptive representation —that is, the proportion of women in EP— but also on the ideological environment in which parliamentarians are embedded (Kantola and Rolandsen-Agustín, 2019). This finding is consistent with recent scholarship highlighting the continued centrality of gender equality as a contested political issue across European institutions, even under conditions of increased female political leadership (Abels et al., 2025).

Beyond ideological affiliation, national political contexts may also contribute to explaining the variation observed across member states. While these factors were not directly examined in the present study, future research could explore their interaction with partisan and institutional dynamics in shaping gendered patterns of political communication.

The sentiment results add a further interpretative layer (Demertzis, 2013). Although women's tweets display slightly higher levels of positivity (Mohammad and Yang, 2011; Nee and De Maio, 2019), the more relevant pattern is that equality-related discourse itself tends to adopt a constructive register across genders. This suggests that, among MEPs, feminism and women's issues are generally communicated through a constructive register rather than through confrontation. In other words, equality discourse appears integrated into mainstream parliamentary communication rather than positioned as a marginal or polarising theme.

Overall, the evidence points to change in emphasis rather than change in structure: gender is associated with expression, while institutional alignments shape connectivity. Gender is associated with differences in thematic focus, communicative style and tonal balance, but it does not become the primary organizing principle of digital interaction. The political logic of the platform remains anchored in established alignments, organised primarily along partisan and national lines, while gender differences operate within those boundaries rather than redefining them.

## Data availability statement

The raw data supporting the conclusions of this article will be made available by the authors, without undue reservation.

## Ethics statement

This study was based exclusively on publicly available data obtained from the public X/Twitter accounts of Members of the European Parliament. It did not involve interaction with human participants, interventions, surveys, interviews, or the collection of private data. Therefore, ethical review and approval was not required. The data were analysed in aggregate form and used solely for academic research purposes.

## Author contributions

AL-U: Writing – original draft, Resources, Investigation, Funding acquisition, Validation, Formal analysis, Writing – review & editing, Methodology, Supervision, Conceptualization, Project administration. SP-F: Writing – review & editing, Conceptualization, Investigation, Supervision, Writing – original draft, Funding acquisition, Resources, Visualization, Validation, Project administration, Formal analysis, Methodology. JO-T: Investigation, Writing – review & editing, Validation, Formal analysis. JM-i: Data curation, Methodology, Validation, Writing – review & editing, Software, Visualization.

## Funding

The author(s) declared that financial support was received for this work and/or its publication. This article is part of the scientific output of the Basque Government's consolidated research group Gureiker (IT1496-22) and of the project "Social Media, Disinformation and AI: Strategies and tools to prevent polarisation and foster democratic values (EHU-G25/05), funded by the University of the Basque Country (EHU)".

## Acknowledgments

The authors are grateful to the editors and reviewers for their insightful comments and constructive feedback throughout the review process.

## Conflict of interest

The author(s) declared that this work was conducted in the absence of any commercial or financial relationships that could be construed as a potential conflict of interest.

## Generative AI statement

The author(s) declared that Generative AI was not used in the creation of this manuscript.

Any alternative text (alt text) provided alongside figures in this article has been generated by Frontiers with the support of artificial intelligence and reasonable efforts have been made to ensure accuracy, including review by the authors wherever possible. If you identify any issues, please contact us.

## Publisher's note

All claims expressed in this article are solely those of the authors and do not necessarily represent those of their affiliated organizations, or those of the publisher, the editors and the reviewers. Any product that may be evaluated in this article, or claim that may be made by its manufacturer, is not guaranteed or endorsed by the publisher.